\documentstyle[twocolumn,prl,graphicx,aps,floats]{revtex}
\begin{document}
\draft

\title{Cross-sections of spallation residues produced in 1$\cdot$A GeV $^{208}$Pb on proton reactions}
\author{ 
W.~Wlaz\l{}o$^{a,f}$,
T.~Enqvist$^b$\footnote{Present address: Oliver Lodge Laboratory, 
University of Liverpool, Liverpool LG9 7ZE, UK.},
P.~Armbruster$^b$, 
J.~Benlliure$^{b,c}$, 
M.~Bernas$^d$, 
A.~Boudard$^a$,
S.~Cz\'ajkowski$^e$,
R.~Legrain$^a$,
S.~Leray$^a$,
B.~Mustapha$^d$,
M.~Pravikoff$^e$,
F.~Rejmund$^{b,d}$,
K.-H.~Schmidt$^b$,
C.~St\'ephan$^d$,
J.~Taieb$^{b,d}$,
L.~Tassan-Got$^d$, 
and~C.~Volant$^a$ }
\address{
$^a$DAPNIA/SPhN  CEA/Saclay, F-91191 Gif-sur-Yvette Cedex, France\\
$^b$GSI, Planckstrasse 1, D-64291 Darmstadt, Germany\\
$^c$University of Santiago de Compostela, 15706 Santiago de Compostela, Spain\\
$^d$IPN Orsay, BP 1, F-91406 Orsay Cedex, France\\
$^e$CEN Bordeaux-Gradignan, F-33175, Gradignan, France\\
$^f$Jagiellonian University, Institute of Physics, ul. Reymonta 4, 30-059 
Krak\'{o}w, Poland}
\date{\today}
\maketitle

\begin{center}
\hspace*{1.6cm}
\begin{minipage}{0.82\textwidth}
\vspace*{-1.2cm}

\begin{abstract}
Spallation residues produced in 1 GeV per nucleon $^{208}$Pb on proton reactions have been studied using the FRagment Separator facility at GSI. Isotopic production cross-sections of elements from $_{61}$Pm to $_{82}$Pb have been 
measured down to 0.1 mb with a high precision. The recoil kinetic energies of the produced fragments were also determined. The obtained cross-sections
agree with  most of the few existing gamma-spectroscopy data. The data are compared with different intranuclear-cascade and evaporation-fission models. Drastic deviations were found for a standard code used in technical applications. 
\end{abstract}
\end{minipage}
\end{center}

Spallation reactions have recently captured an increasing interest due to their technical applications as intense neutron sources for accelerator-driven sub-critical reactors \cite{LasVegas} or spallation neutron sources \cite{Carp}. The design of an accelerator-driven system (ADS) requires precise knowledge of nuclide production cross-sections in order to be able to predict the amount of radioactive isotopes produced inside the spallation target. Indeed, short-lived isotopes may be responsible for maintenance problems and long-lived ones will increase the long term radiotoxicity of the system. Recoil kinetic energies of the fragments are important for studies of radiation damages in the structure materials or in the case of a solid target. Data concerning lead are particularly important since in most of the ADS concepts actually discussed, lead or lead-bismuth alloy is considered as the preferred material of the spallation-target.

The present experiment, using inverse kinematics is able to supply the identification of all the isotopes produced in spallation reactions and information on their recoil velocity. Moreover, the data represent a crucial benchmark for the existing spallation models used in the ADS technology. The precision of these models to estimate residue production cross-sections is still far from the performance required for technical applications, as it was shown in Ref.\cite{Mic97}. This can be mostly ascribed to the lack of complete distributions of all produced isotopes to constrain the models. The available data were generally obtained by chemistry or gamma spectroscopy \cite{Wolf56,Mic2,Mic99} which give access mostly to cumulative yields produced after long chains of decaying isotopes.

In this Letter we report on complete isotopical production cross-sections for heavy fragments produced in spallation of $^{208}$Pb on proton at 1$\cdot$A GeV, down to 0.1 mb with a high precision. The kinematic properties of\linebreak[4]
\vspace*{2.65cm}

\noindent
the residues are also studied. The cross-sections of lighter isotopes produced by fission will be presented in a forthcoming publication.

The experimental method and the analysis procedure have been developed and applied in previous experiments \cite{jong,timou,ben99}. The primary beam of 1$\cdot$A GeV $^{208}$Pb was delivered by the heavy-ion synchrotron SIS at GSI, Darmstadt. The proton target was composed of 87.3 mg/cm$^2$ liquid hydrogen \cite{hydro} enclosed between thin titanium foils of a total thickness of 36 mg/cm$^2$. The primary-beam intensity was continuously monitored by a beam-intensity monitor (SEETRAM) based on secondary-electron emission. In order to subtract the contribution of the target windows from the measured reaction rate, measurements were repeated with the empty target. Heavy residues produced in the target were all strongly forward focused due to the inverse reaction kinematics. They were identified using the FRagment Separator (FRS) \cite{Gei92}. 

The FRS is a two-stage magnetic spectrometer with a dispersive intermediate image plane (S$_2$) and an achromatic final image plane (S$_4$) with momentum acceptance of 3 \% and angular acceptance of 14.4 mrad around the beam axis. Two position-sensitive plastic scintillators placed at S$_2$ and S$_4$, respectively, provided the magnetic-rigidity (B$\rho$) and time-of-flight measurements, which allowed to determine the mass-over-charge ratio of the particles. In the analysis, totally stripped residues were considered only. In the case of residues with the highest nuclear charges (above $_{65}$Tb) an achromatic degrader (5.3 g/cm$^2$ to 5.9 g/cm$^2$ of aluminum) was placed at S$_2$ to obtain a better Z resolution. The elements below terbium were identified from an energy-loss measurement in an ionisation chamber (MUSIC). 
The velocity of the identified residue was determined at S$_2$ from the B$\rho$ value and transformed into the frame of the beam in the middle of the target taking into account the appropriate energy loss. About 100 different values of the magnetic field were used in steps of about  2 \% in order to cover all the produced residues and to construct the full velocity distribution of each residue.

\begin{figure}[bhtp]
\centering
\includegraphics[width=7.5cm,angle=0]{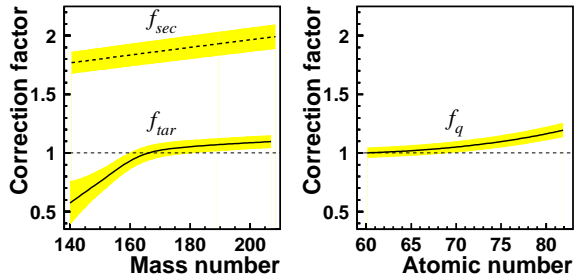}
\vspace*{0.25cm}
\caption{
Main correction factors applied to the data. Left panel: factors due to secondary reactions in all layers of matter of the FRS, including the degrader, $f_{sec}$
(dashed line), and secondary reactions and beam attenuation inside the hydrogen target, $f_{tar}$ (solid line), versus fragment mass number. Right panel: factor corresponding to not fully stripped ions, $f_{q}$, versus fragment atomic number. The gray area around each line denotes the associated systematic error.  }
\label{corect}
\end{figure}

The measured counting rates $N$, attributed to a specific isotope, were normalized to the number of projectiles $N_p$ recorded with the beam-intensity monitor and to the number of target atoms per area $n_t$. Then, the production cross-sections $\sigma_{prod}$ are calculated by applying several corrections 
according to the following equation: 
\begin{equation}
\sigma_{prod}=\frac{N}{n_t 
\cdot N_p} \cdot f_\tau \cdot f_\epsilon \cdot f_{tr}
\cdot f_{q}\cdot f_{tar} \cdot f_{sec}.  
\end{equation}

The used correction factors arise from: the measured dead-time of the data-acquisition system ($f_\tau$), the efficiency of the detection system ($f_\epsilon$), the loss of fragments due to the fragment-separator transmission ($f_{tr}$), the loss of fragments due to incompletely stripped ions ($f_{q}$), 
the influence of beam attenuation and secondary reactions in the liquid-hydrogen target ($f_{tar}$), and in the other layers of matter inside the fragment separator ($f_{sec}$). The $f_\tau$, $f_\epsilon$ and  $f_{tr}$ correction factors are directly deduced from the experiment with high precision. The dead time was measured by the acquisition system and kept below 30 \%. The efficiency of all detectors was estimated directly from the obtained data to be higher than 98 \%. Since the full velocity distribution is constructed for each isotope from the data of different field settings, transmission losses are negligeable in the present experiment.

The corrections due to incompletely stripped ions ($f_{q}$) and secondary reactions ($f_{tar}$ and $f_{sec}$)  depend on the fragment type. They are displayed with their associated uncertainties in Fig. \ref{corect}. $f_{q}$ represents the counting loss due to incompletely stripped ions. It is significant only in case of the fragments with a high nuclear charge. For several isotopes the ratio between fully and incompletely stripped ions was determined. It made possible to estimate the loss in counting rate due to the fraction of incompletely stripped ions for all isotopes. $f_{sec}$ corresponds to the loss of residues through secondary reactions in the thick aluminum 
degrader and other layers of matter in the beam line. It was calculated using two different formulas for the total reaction cross-sections, developed by Karol \cite{Karol} and Benesh \cite{Benes}. The results agreed within 5 \%. $f_{sec}$ varies from 2 to 1.8 with decreasing mass. However for Gd, Eu, Sm and Pm 
whose cross-sections were collected without the degrader, it is not higher than  1.13. Secondary reactions inside the hydrogen target also lead to a reduction of the counting rates of the heaviest isotopes, but on the other hand produce more lighter isotopes. The corresponding correction factor, $f_{tar}$, was estimated from reaction rates obtained in the present experiment using a deconvolution method. The total reaction cross-section formula of Benesh {\it et al}.  \cite{Benes} utilized in these calculations was adjusted to the experimental data from the Barashenkow compilation \cite{bara}. 

\begin{figure}[tbh]
\begin{minipage}{1.0\textwidth}
\centering
\includegraphics[width=14.cm,angle=0]{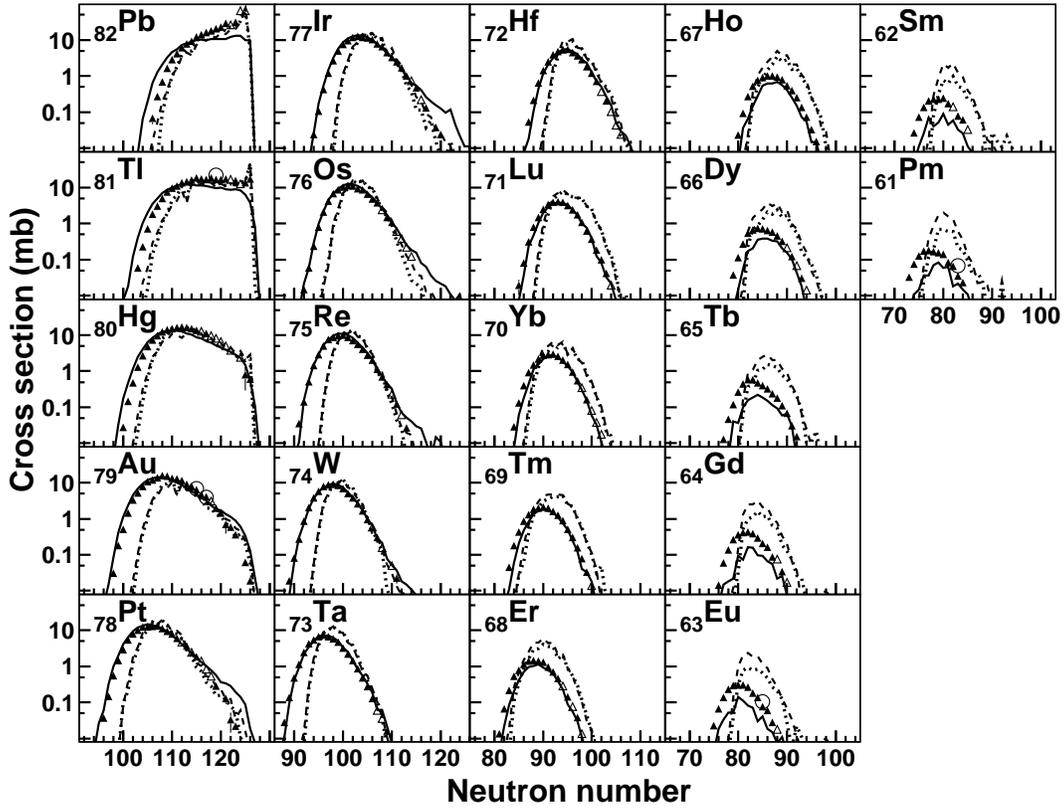}
\vspace*{0.25cm}
\caption{
Isotopic production cross-sections of elements between Z=82 and 61, in the reaction of 1$\cdot$A GeV $^{208}$Pb on hydrogen, versus neutron number. Stable (resp. radioactive) isotopes are marked by open (resp. full) triangles. Gamma-spectroscopy data regarding shielded isotopes from \protect\cite{Mic99} are plotted as open circles. The solid, dashed and dotted curves were calculated with the Cugnon-Schmidt \protect\cite{cugnon,abla}, Bertini~\protect\cite{ber}-Dresner~\protect\cite{dre,atch} and Isabel\protect\cite{isa}-Dresner models, respectively.}
\label{isotopic}
\end{minipage}
\end{figure}

All uncertainties of the used corrections lead to a final systematic error of 9 \% to 23 \% for Z from 82 to 61. The measured production cross-sections of the spallation residues in the reaction of 1$\cdot$A  GeV $^{208}$Pb with protons are plotted as isotopic distributions in Fig. \ref{isotopic}. Most of the presented distributions exhibit a Gaussian-like shape where the neutron-proton evaporation competition determines the position of the maximum. The most significant deviations from this shape occur for the neutron-rich fragments 
with masses close to that of the projectile. In the case of these residues, 
one and a few neutron-removal channels from low excited nuclei created mainly 
in peripheral collisions are responsible for the increased production cross-sections. Most of the produced isotopes populate a corridor, between the valley of stability and the proton drip line due to the fact that the  excited heavy prefragment evaporates mainly neutrons. 

\begin{table}[bhtp]
\caption{
Comparison of the cross-sections from the present work ($\sigma_{FRS}$) with 
those obtained by gamma spectroscopy ($\sigma_{RC}$) by Gloris {\em et al.} 
\protect\cite{Mic99}. The  $\sigma_{RC}$ values are given with total error. The  $\sigma_{FRS}$ cross-sections are with statistical and total (in parentheses) error. In the fourth column the systematic uncertainty for each  $\sigma_{FRS}$ is given.
}
\begin{tabular}{lccc}  
Isotope    &   $\sigma_{RC}$ (mb) & $\sigma_{FRS}$ (mb) & Systematic error (\%) \\ \hline
$^{200}$Tl &   22.3$\pm$6.1    &  17.0$\pm$0.4(1.6)    & 9\\   
$^{196}$Au &   3.88$\pm$0.47   &  4.0$\pm$0.1(0.4)   & 9\\
$^{194}$Au &   6.85$\pm$0.92   &  6.3$\pm$0.2(0.6)   & 9\\ 
$^{148}$Eu &   0.104$\pm$0.04  &  0.075$\pm$0.005(0.010) & 12\\ 
$^{144}$Pm &   0.068$\pm$0.013 &  0.036$\pm$0.003(0.006) & 15\\ 
\end{tabular}
\label{mich}
\end{table}

In Fig. \ref{isotopic} the cross-sections obtained by gamma spectroscopy  
\cite{Mic99} are also shown. To compare with our data, we have chosen only 
isotopes shielded by long-lived or stable precursor in the decay chain. A more detailed comparison is presented in Table \ref{mich}. In the case of $^{196}$Au, the cross-section is the sum of the production of the ground and the isomeric states. The data agree within their error bars, except for the isotope with the lowest cross-section, $^{144}$Pm.

\begin{figure}[hbtp]
\centering
\includegraphics[width=8.0cm,angle=0]{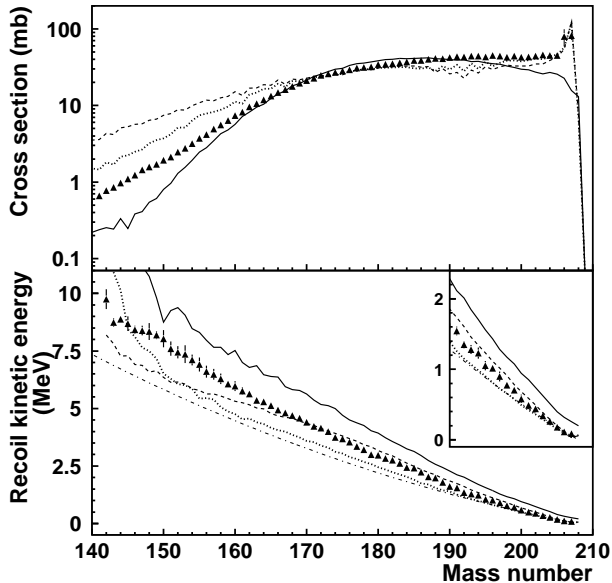}
\vspace*{0.25cm}
\caption{
Mass distribution (upper panel) and recoil kinetic energy (bottom panel) 
of the residues produced in 1$\cdot$A GeV $^{208}$Pb on hydrogen reactions 
(triangles) versus mass number, compared with the Cugnon-Schmidt (solid line), Bertini-Dresner (dashed line) and Isabel-Dresner (dotted line) models. The dash-dotted line shows the recoil kinetic energies expected from the Morrissey systematics \protect\cite{moris}.}
\label{mass}
\end{figure}

Spallation reactions are generally modeled as a two-step process. In the first step, the nucleon-nucleon collisions inside the nucleus induce the loss of a few nucleons and lead to the formation of an excited prefragment. This process can be described by the intranuclear cascade model (INC) sometimes including a pre-equilibrium emission. In the second step, the prefragment deexcites by evaporation of light particles or by fission. Calculations performed with different INC plus evaporation-fission models are shown together with our results in Figs.~\ref{isotopic} and \ref{mass}. The first two calculations were done with the commonly used LAHET Code System (version 2.7 with default options) from Los Alamos~\cite{lahet} using either the Bertini~\cite{ber} plus pre-equilibrium (dashed line) or Isabel~\cite{isa} (dotted line) INC models followed by the Dresner evaporation-fission model~\cite{dre,atch}. The shapes of the isotopic distributions obtained with both INC models are very similar and differ significantly from the experimental ones: they are shifted with respect to the experimental ones towards the neutron-rich side. This can be ascribed\linebreak[4]
\vspace*{12.7cm}
 
\noindent
 to the fact that the prediction of the neutron-proton evaporation competition in the Dresner code is not satisfying. The magnitude of the measured and calculated cross-sections is also quite different, especially in the case of the lighter elements. This effect is better visible on the mass distribution (Fig. \ref{mass}, upper panel) and more marked with the Bertini model which overpredicts largely the production of light isotopes. This discrepancy of the  Bertini model is due to a distribution of excitation energies ($E^{\star}$) of the prefragments extending to too high values, which results in evaporating more particles and finally producing lighter nuclides. This problem of a too high $E^{\star}$ at the end of the Bertini INC model was already noticed in a comparison with neutron double-differential cross-section measurements \cite{saturn} although, here, the use of the pre-equilibrium option has led to somewhat smaller $E^{\star}$. On the other hand, in a region very close to the projectile mass, both Bertini and Isabel calculations are in good agreement with the data.

The third calculation (solid line in Fig. \ref{isotopic} and \ref{mass}) was performed with the version INCL3 of the Cugnon model \cite{cugnon} combined with a model elaborated by Schmidt {\em et al.} \cite{abla}. This calculation reproduces much better than the former ones the shape of the experimental isotopic distributions. This comes mainly from a better description of the neutron-proton competition in the Schmidt than in the Dresner evaporation model, since the $E^{\star}$ distribution at the end of the INC stage is similar in the Isabel and Cugnon models (except for very small $E^{\star}$). For elements from $_{76}$Os to $_{79}$Au the code predicts shoulders on the neutron-rich side of the isotopic distributions. We attribute these  to the statistical treatment of the Pauli blocking in the Cugnon model which improves significantly the excitation-energy distribution \cite{saturn} in general but also leads to a few prefragments with unrealistically low excitation energies.
This problem is already partly cured in INCL3 compared to the previous version \cite{cugnon}. The magnitude of the cross-sections is not always reproduced, the calculation under-predicting the production of the light isotopes. 
Besides, the main defect of this calculation is the underproduction of isotopes very close to the projectile, which represent an important part of the total cross-section. This is ascribed to the sharp surface approximation in the Cugnon model which leads to a bad description of the most peripheral reactions. These defects result in a poor prediction of the mass distribution. 

 The velocity distribution of each residue was also determined, from which it was possible to infer information about the recoil kinetic energy in the projectile system. In the bottom part of Fig. \ref{mass}, the average recoil kinetic energy of the fragments is shown as a function of their mass number. The systematic uncertainty of the obtained values (not shown in the picture) varies from about 8 \% to 30 \% for A from 140 to 208. Calculations performed with the Cugnon-Schmidt (solid line), Bertini-Dresner (dashed line) and Isabel-Dresner (dotted line) codes are also shown, together with an empirical parameterization (dash-dotted line) describing the average longitudinal momentum transfer distributions derived by Morrissey from a large compilation of experimental data \cite{moris}. The Cugnon-Schmidt code predicts recoil energies up to 35 \% higher than the experimental ones while the three other calculations underestimate the data for large mass losses. 

In conclusion, the fragment-separator facility at GSI has been used to determine, for the first time, the production cross-sections and momentum distributions of 446 isotopes from spallation reactions of $1\cdot$A GeV $^{208}$Pb with protons. The results agree with most of the few cross-sections previously measured by gamma-spectroscopy. Calculations using different models 
have been performed. Although none of them provide a detailed description
of the data, the new Cugnon-Schmidt code gives clear improvements.


\begin{thebibliography}{99}

\vspace*{-1.8cm}
\bibitem{LasVegas} {\it Proceedings of the Int. Conf.
        on Accele\-rator-Driven Transmutation Technologies and
        Applications}, Las Vegas, 1994, edited by E.D. Arthur, A.
        Rodrigues, and S.O. Schriber (AIP Press, Woodbury, NY, 1995).
%
\bibitem{Carp} J.M. Carpenter, Nucl. Instr. and Meth. {\bf 145}, 91 (1977).
%    
\bibitem{Mic97} R. Michel and P. Nagel, {\it International Codes and Model 
        Intercomparison for Intermediate Energy Activation Yields}
        OECD/NEA, Paris, 1997. 
%
\bibitem{Wolf56} R. Wolfgang et al., Phys. Rev.  103 (1956) no. 2. 
%
\bibitem{Mic2} R. Michel et al., Nucl. Instrum. Methods B129 (1997) 153. 
%
\bibitem{Mic99} M. Gloris et al., accepted for publication in Nucl. Instrum. Methods (2000).
%
\bibitem{jong} M. de Jong et al., Nucl. Phys. A 628 (1998) 479.
%
\bibitem{timou} T. Enqvist et al., Nucl. Phys. A 658 (1999) 47.
%
\bibitem{ben99} J. Benlliure et al., Nucl. Phys. A 660 (1999) 87.
%
\bibitem{hydro} P. Chesny et al.,GSI Ann. Rep. (1996) 190.
%
\bibitem{Gei92} H. Geissel et al., Nucl. Instrum. Methods B70 (1992) 
286.
%
\bibitem{Karol} P.J. Karol, Phys. Rev. C11 (1975) 1203.
%
\bibitem{Benes} C.J. Benesh et al., Phys. Rev. C40 (1989) 1198.
%
\bibitem{bara} B. C. Barashenkov, 
Cross Sections of Interactions of Particle and Nuclei with Nuclei, 
JINR, Dubna, 1993. 
%
\bibitem{lahet} R. E. Prael and H. Lichtenstein, Los Alamos-UR-89-3014 (1989). 
%
\bibitem{ber} H.W. Bertini, Phys. Rev. 188 (1969) 1711.
%
\bibitem{isa} Y. Yariv and Z. Frankel, Phys. Rev. C20 (1979) 2227.
%
\bibitem{dre} L. Dresner, report ORNL-TM-196, Oak Ridge National Laboratory (1962).
%
\bibitem{atch} F. Atchison, Targets for Neutron
Beam Spallation Sources, J\"{u}l-Conf-34, Kernforschungsanlage J\"{u}lich GmbH 
(1980).
%
\bibitem{cugnon} J. Cugnon, C. Volant, S. Vuillier, Nucl. Phys. A620 (1997) 
457.
%
\bibitem{abla} A. R. Junghans et al., Nucl. Phys. A 629 (1998) 635.
%
\bibitem{saturn} X. Ledoux et al., Phys. Rev. Lett. 82 (1999) 22.
%
\bibitem{moris} D.J. Morrissey, Phys. Rev. C 39  (1989) 460.
%
\end{thebibliography}
\end{document}